\documentclass{appolb}
\usepackage{graphicx}
\usepackage[T1]{fontenc}
\begin{document}
\title{Present status and future prospects of neutrino oscillation experiments%
\thanks{Presented at  ``Matter to the deepest: Recent developments in physics of fundamental interactions, XLV international conference of theoretical physics (MTTD 2023)", Ustroń, Poland, Sep 17 – 22, 2023}% 
}
\author{Monojit Ghosh
\thanks{mghosh@irb.hr}
\address{Center of Excellence for Advanced Materials and Sensing Devices, Ru{\dj}er Bo\v{s}kovi\'c Institute, 10000 Zagreb, Croatia}
\\
}
\maketitle
\begin{abstract}
In this proceeding we discuss the status of the currently running experiments and the capability of the future proposed experiments to study neutrino oscillation. In particular, we discuss the current results of the accelerator-based long-baseline experiments in the standard three-flavour scenario and for a scenario where one assumes the existence of a light sterile neutrino at the eV scale in addition to the three active neutrinos.  Further, we also discuss the capability of the future long-baseline experiments to study these scenarios. 

\end{abstract}
  
\section{Introduction}
In the quantum mechanical interference phenomenon of neutrino oscillation, neutrinos undergo transitions from one flavour to another \cite{Akhmedov:1999uz}. The oscillations of the neutrinos occur over the macroscopic distance and time. This happens because the neutrino flavour states and the neutrino mass states are not same and they are related by the unitary PMNS matrix. Apart from the oscillations in the standard three flavour framework \cite{Esteban:2020cvm}, there are many new physics scenarios, which can affect the neutrino oscillations \cite{Denton:2022pxt}. In this proceeding, we will review the present status and the future prospects of the neutrino oscillation experiments in the standard three-flavour framework and in the presence of a light sterile neutrino with mass in the eV range i.e., the 3+1 case \cite{Acero:2022wqg}. Neutrino oscillation experiments are solar based, atmospheric based, reactor based and accelerator based \cite{Ghosh:2015oqc}. In this proceeding, we will discuss the current status and future capabilities of the accelerator based neutrino oscillation experiments \cite{Agarwalla:2014fva}. 

This proceeding is oraganized as follows. In section~\ref{3nu} we will discuss the neutrino oscillation in the standard three-flavour scenario and in section~\ref{3+1nu} we will discuss the same in the 3+1 scenario. In each of these sections, we first discuss the current status and then the future prospects. In addition, section~\ref{oth} demonstrates briefly the various new physics scenarios apart from the 3+1 case, which are being studied in the context of different neutrino oscillation experiments. Finally in section~\ref{sum} we summarize and give our concluding remarks.

\section{Oscillations in the standard 3 flavour scenario}
\label{3nu}

In the standard three-flavour scenario the PMNS matrix, which connects the flavour sates and the mass states of the neutrinos, can be expressed as \cite{ParticleDataGroup:2022pth}:
\begin{eqnarray}
U^{3\nu}_{\rm PMNS} = R_{23}(\theta_{23}) R_{13} (\theta_{13}, \delta_{\rm CP}) R_{12} (\theta_{12}),
\label{pmns3}
\end{eqnarray}
where $R_{ij}$ is the $3 \times 3$ rotation matrices in the $i$-$j$ plane. From Eq.~\ref{pmns3}, it is evident that the mixing matrix is a function of three mixing angles i.e., $\theta_{12}$, $\theta_{23}$ and $\theta_{13}$ and one Dirac type CP phase i.e., $\delta_{\rm CP}$.  Apart from these, neutrino oscillation in three-flavour also depends upon two mass squared differences i.e., $\Delta m^2_{21} = m_2^2 - m_1^2$ and $\Delta m^2_{31} = m_3^2 - m_1^2$, where $m_1$, $m_2$ and $m_3$ are the mass of the active neutrinos. Measurements of the parameters $\theta_{12}$ and $\Delta m^2_{21}$ come from the solar neutrino experiments whereas the measurements of  $\theta_{23}$ and $\Delta m^2_{31}$ come from the atmospheric neutrino experiments. The reactor neutrino experiments are sensitive to the parameter $\theta_{13}$ and the accelerator-based neutrino experiments are sensitive to the parameters $\delta_{\rm CP}$, sign of $\Delta m^2_{31}$ and octant of $\theta_{23}$. As mentioned in the introduction, here we focus on the accelerator-based experiments, which are also known as long-baseline experiments. 

\subsection{Current status}

\begin{figure}[htb]
\centerline{
\includegraphics[width=10cm]{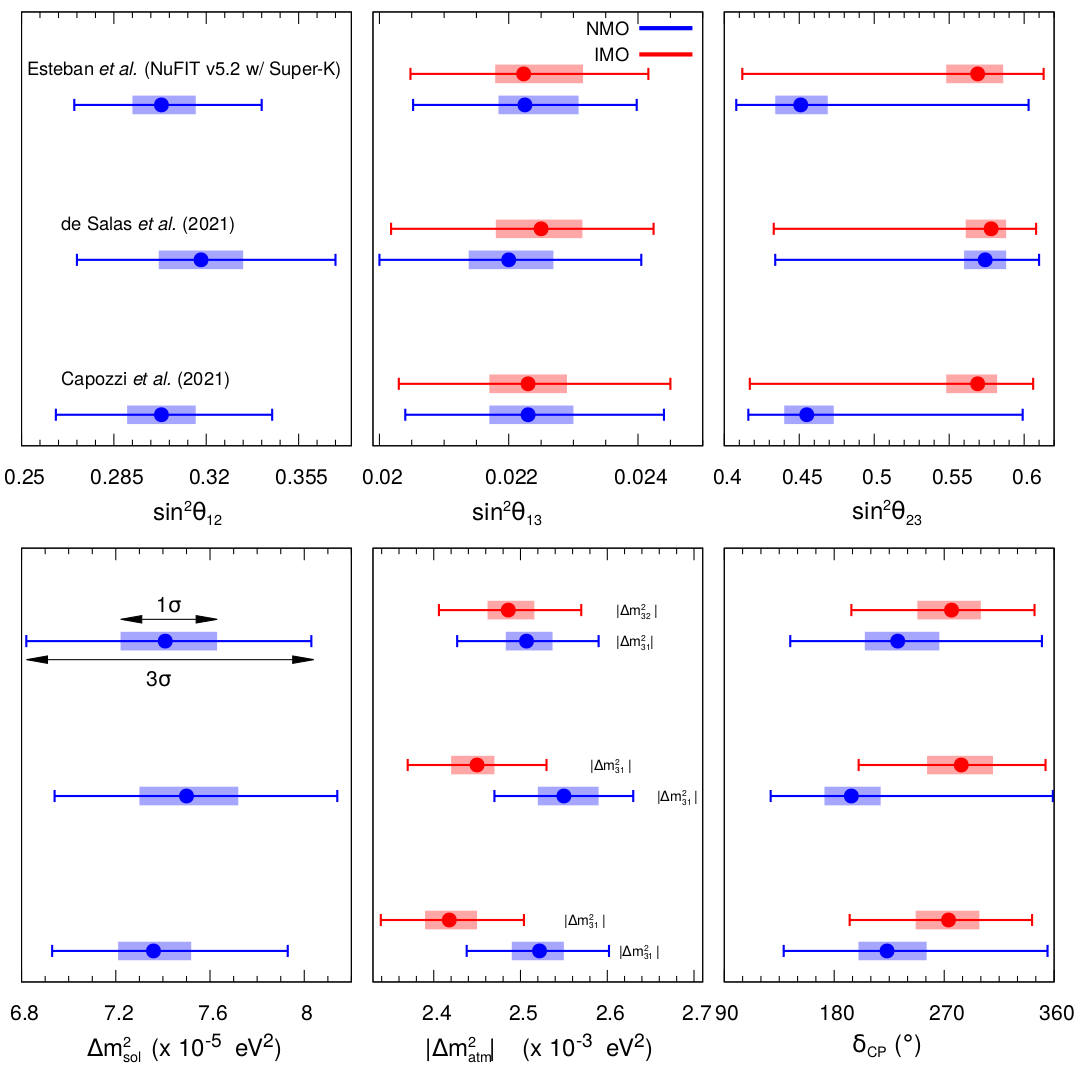}}
\caption{Current status of the neutrino oscillation parameters in the standard three flavour scenario. The phrases ``NMO'' stands for normal ordering of the neutrino masses and ``IMO'' stands for inverted ordering of the neutrino masses. The figure is taken from Ref.~\cite{Agarwalla:2021bzs}.}
\label{fig1}
\end{figure}

The current status of the above mentioned neutrino oscillation parameters are shown in Fig.~\ref{fig1} \cite{Agarwalla:2021bzs}. Currently there are three groups involved in performing the global analysis of the world neutrino data. These groups are Nufit group \cite{Esteban:2020cvm}, Valencia group \cite{deSalas:2020pgw} and Bari group \cite{Capozzi:2021fjo}.  Their recent results are compiled in Fig.~\ref{fig1}.  At present the unknowns in this sector are: (i) the ordering of the neutrino masses, which can be either normal or inverted, (ii) octant of the mixing angle $\theta_{23}$, which can be higher of lower and (iii) the value of $\delta_{\rm CP}$. Further, the precision of the atmospheric parameters i.e., $\theta_{23}$ and $\Delta m^2_{31}$  needs to be improved further. From the figure we see that the results from the three groups concerning the values of the oscillation parameters more or less agree with each other except the octant of $\theta_{23}$. For Nufit and Bari group, lower octant is favoured for normal ordering whereas Valencia group finds higher octant to be the preferred octant. 
%uncomment the following lines to place a figure

\begin{figure}[htb]
\centerline{
\includegraphics[width=10cm]{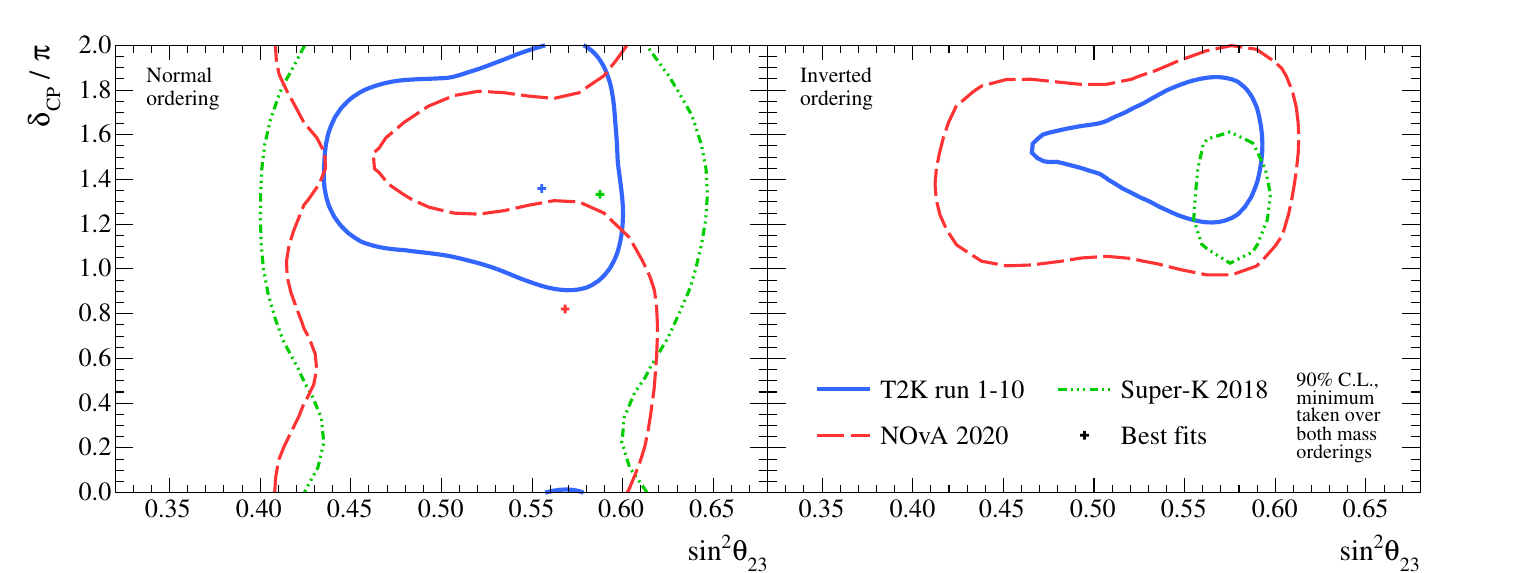}
\hspace{-0.35 in}
\includegraphics[width=5cm]{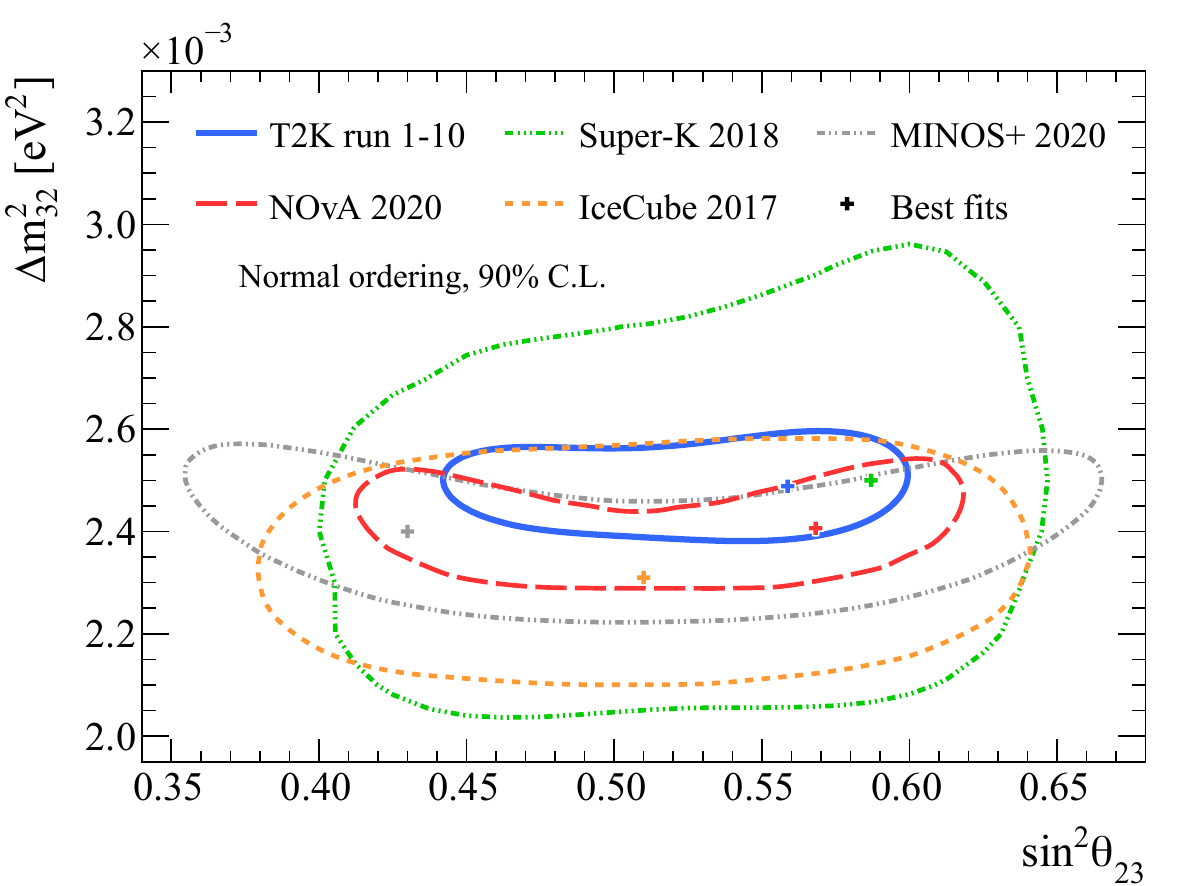}}
\caption{Recent results from T2K and NO$\nu$A regarding the measurement of mass ordering, octant of $\theta_{23}$, $\delta_{\rm CP}$ and the precision of the atmopsheric parameters. The figure is taken from Ref.~\cite{T2K:2023smv}.}
\label{fig2}
\end{figure}

The most recent results regarding the unknowns in the standard three-flavour scenario come from the T2K \cite{T2K:2023smv} experiment in Japan and NO$\nu$A \cite{NOvA:2023iam} experiment in USA. Their recent results are shown in Fig.~\ref{fig2}. This figure shows that normal ordering of the neutrino masses is favoured over the inverted ordering. Higher octant of $\theta_{23}$ is preferred over the lower octant. Regarding $\delta_{\rm CP}$, there is a mismatch between these experiments. The best-fit of T2K gives around $-90^\circ$ whereas the best-fit of NO$\nu$A is around $180^\circ$. Regarding the precision of $\theta_{23}$ and $\Delta m^2_{31}$, we see that the sensitivity at present has improved as compared to results in 2018 \cite{T2K:2023smv}.

\subsection{Future Prospects}

Currently, there are three proposed long-baseline experiments, which are expected to measure the remaining unknowns in the neutrino oscillation with very high confidence level. These experiments are T2HK \cite{Hyper-Kamiokande:2016srs} in Japan, DUNE \cite{DUNE:2020ypp} in USA and ESSnuSB \cite{Alekou:2022emd} in Sweden. In the first phase, T2HK in Japan, which is basically the upgrade of the T2K experiment, will use a water Cherenkov detector of volume 187 kt located at a distance of 295 km from the neutrino source at J-PARC to study the oscillation of the muon neutrinos. For DUNE, a LAr time projection chamber detector of 40 kt will be located at a distance of 1300 km from the neutrino source at Fermilab, whereas ESSnuSB will use a water Cherenkov far detector of mass 540 kt  located at a distance of 360 km from the neutrino source at the ESS facility.  The experiments T2HK and DUNE will study the neutrino oscillation at the first oscillation maximum whereas the experiment ESSnuSB will probe the second oscillation maximum, specifically the furthest part of the first oscillation maximum and the beginning of the second oscillation maximum, which is also known as the first oscillation minimum.

\begin{figure}[htb]
\centerline{
\includegraphics[width=3.8cm]{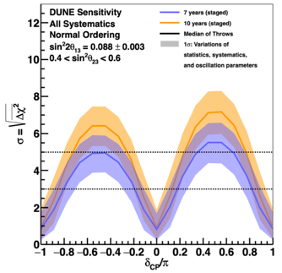}
\includegraphics[width=4.7cm]{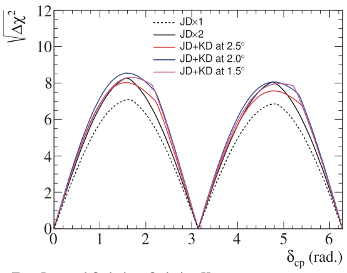}
\includegraphics[width=5.4cm]{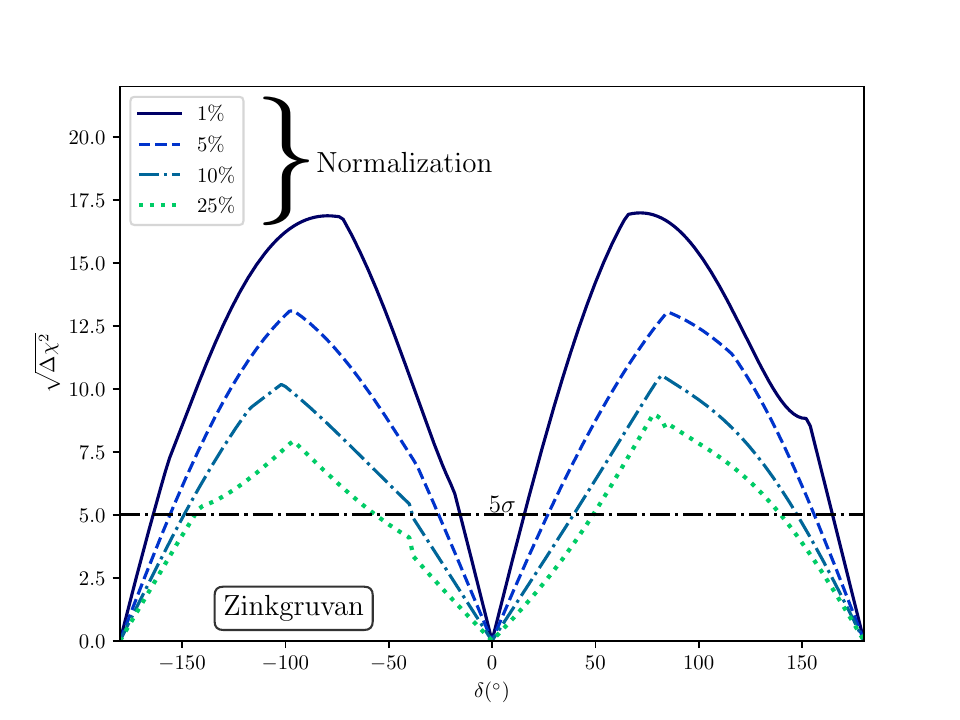}}
\centerline{
\includegraphics[width=3.9cm]{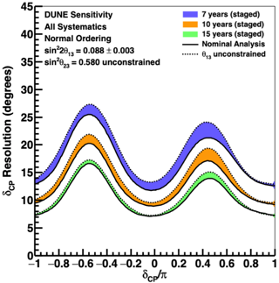}             
\includegraphics[width=5.1cm]{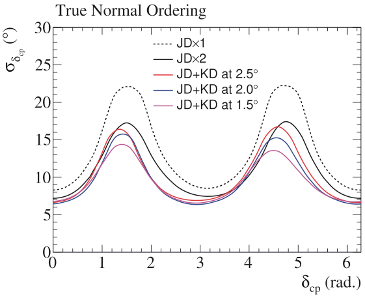}
\includegraphics[width=5.0cm]{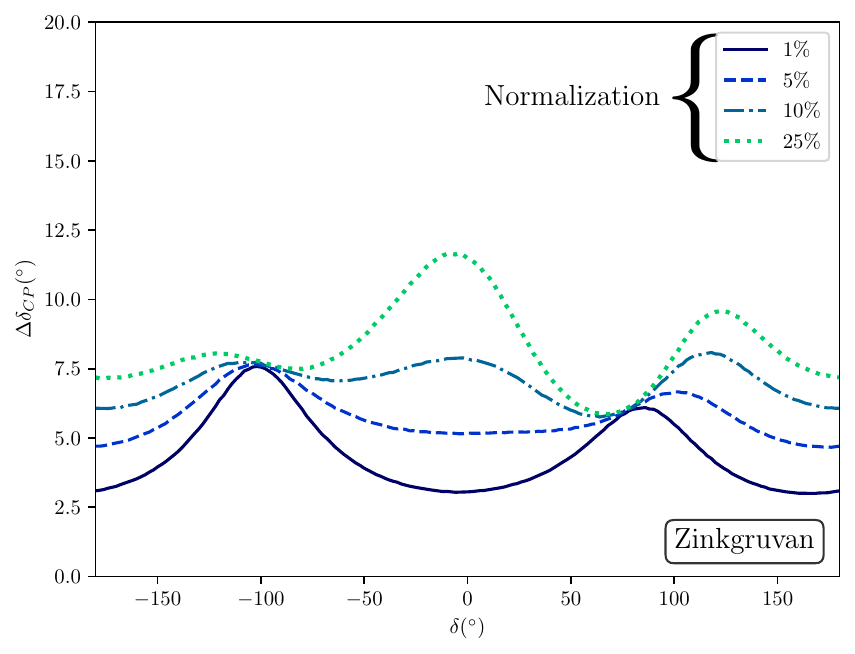}}
\centerline{
\includegraphics[width=4.6cm]{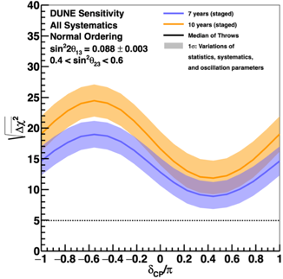}
\includegraphics[width=4.8cm]{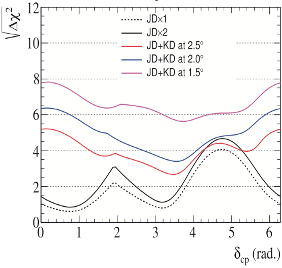}
\includegraphics[width=4.6cm]{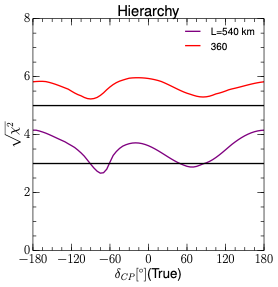}}
\caption{CP violation sensitivity (top row), CP precision sensitivity (middle row) and mass ordering sensitivity (bottom row) sensitivity. Left column is for DUNE, middle panel is for T2HK and the right column is for ESSnuSB. For T2HK, the JD $\times$ 1 curves correspond to the detector volume of 187 kt.  For ESSnuSB, different curves are for different values of systematic uncertainties corresponding to normalization error. The figures are taken from Refs.~\cite{Hyper-Kamiokande:2016srs,DUNE:2020ypp, Alekou:2022emd,ESSnuSB:2021azq}.}
\label{fig3}
\end{figure}

The sensitivities of T2HK, DUNE and ESSnuSB are presented in Fig.~\ref{fig3}. From the figure we see that the CP sensitivity is best for the ESSnuSB experiment. This arises from the large CP sensitivity at the second oscillation maximum. The sensitivity of T2HK is also quite high, thanks to of large statistics, which stems from its shorter baseline. For neutrino mass-ordering the best sensitivity comes from the DUNE experiment. This is because of the longer baseline and larger matter effect. 

\section{Oscillations in the 3+1 scenario}
\label{3+1nu}

Sterile neutrinos are SU(2) singlets and they do not have any interactions with the Standard Model particles. However, because the active neutrinos can oscillate into sterile neutrinos, these latter can be probed with the neutrino oscillation experiments. In the 3+1 scenario, the PMNS matrix can be written as:
\begin{eqnarray}
U^{4\nu}_{\rm PMNS} = R_{34}(\theta_{34}, \delta_{34}) R_{24} (\theta_{24}, \delta_{\rm 24}) R_{14} (\theta_{14})U^{4\nu}_{\rm PMNS}.
\label{pmns4}
\end{eqnarray}
Here $R_{ij}$ is the $4 \times 4$ rotation matrices in the $i$-$j$ plane. In this case we have three additional mixing angles i.e., $\theta_{14}$, $\theta_{24}$ and $\theta_{34}$, two additional Dirac type phases i.e., $\delta_{24}$ and $\delta_{34}$ and one additional mass squared difference i.e., $\Delta m^2_{41} = m_4^2 - m_1^2$ where $m_4$ is the mass of the sterile neutrino. 

\subsection{Current status}

The main evidence of sterile neutrinos comes from the LSND~\cite{LSND:2001aii} experiment where a significant excess of events have been seen over the known backgrounds. This was later confirmed by the MiniBooNE~\cite{MiniBooNE:2020pnu} experiment. The recent results of MicroBooNE~\cite{MicroBooNE:2021tya} have sparked the discussion of the light sterile neutrinos once again. According to MicroBooNE, there is no excess of $\nu_e$ events coming from the $\nu_\mu$ beam. However, a combined fit of MiniBooNE and MicroBooNE shows that 3+1 model is still allowed at a significant confidence level~\cite{MiniBooNE:2022emn}. In addition,  data from currently running experiments ICARUS and SBND are also expected to shed some light on the sterile neutrino situation \cite{Bonesini:2022pwy}. 

\subsection{Future Prospects}

\begin{figure}[htb]
\centerline{
\includegraphics[width=6cm]{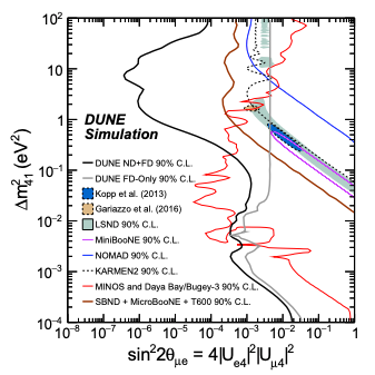}
\includegraphics[width=8cm]{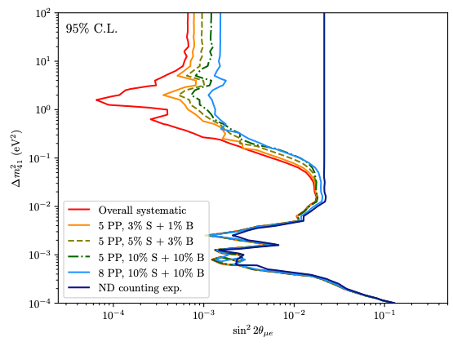}}
\caption{Sensitivity of DUNE (lef panel) and ESSnuSB (right panel) to exclude the sterile mixing parameter space. For ESSnuSB, the curves are for FD+ND with different options for systematic uncertainties. The figures are taken from Refs.~\cite{Ghosh:2019zvl,DUNE:2020fgq}.}
\label{fig4}
\end{figure}

In Fig.~\ref{fig4}, we have shown the sensitivity of the future experiments DUNE and ESSnuSB for both far detector (FD) and near detector (ND) to put upper bounds on the sterile neutrino mixing parameters.  This shows that a stringent bound on the sterile neutrinos can be obtained when one combines the near and the far detector data.  Additionally, for ESSnuSB there is a proposal \cite{ESSnuSB:2023ogw} to build a low energy muon storage ring (LEnuSTORM) similar to the nuSTORM \cite{nuSTORM:2022div} project and a low energy monitored neutrino beam line (LEMNB), inspired by the ENUBET project \cite{Longhin:2022tkk}. With these facilities it will be possible to study sterile neutrinos at the near detector, which will further improve the sensitivity of the ESSnuSB experiment. 

\section{Other new physics scenarios}
\label{oth}

Note that in this proceeding we discussed the current status and future prospects for the sterile neutrinos, which is one of the new physics scenarios that can be probed in the neutrino oscillation experiments. Apart from this, there are several other new physics scenarios, which are currently being studied for the currently running and the future proposed experiments. Some of the examples of these new physics scenarios are: non standard neutrino interactions \cite{ESSnuSB:2023lbg}, violation of Lorentz invariance \cite{Raikwal:2023lzk}, neutrino decay \cite{Choubey:2020dhw} etc. 

\section{Summary and conclusion}
\label{sum}

In this proceeding we discussed the present status and future prospects of the neutrino oscillation experiments. In particular, we discussed the present results and future capabilities of accelerator based long-baseline experiments for the standard three-flavour scenario and for the sterile neutrinos. 

In the standard three-flavour scenario, we show that the global analysis of of the world neutrino data by the different groups are more or less consistent with each other. The recent results from T2K and NO$\nu$A show a mild preference towards the normal ordering of the neutrino masses and the upper octant of the atmospheric mixing angle $\theta_{23}$. Regarding $\delta_{\rm CP}$ there exists a mild tension between the data of T2K and NO$\nu$A. We have also shown that the current precision of the atmospheric mixing parameters $\theta_{23}$ and $\Delta m^2_{31}$ have improved as compared to the results of 2018. For future experiments, we have shown that ESSnuSB and T2HK have excellent capability to measure the phase $\delta_{\rm CP}$ in terms of both CP violation and CP precision measurements whereas DUNE will have the best possible sensitivity towards the measurement of neutrino mass ordering. 

For the search of a light sterile neutrino, we showed that the current status is very intriguing. While the current data from MicroBooNE is against the existence of a light sterile neutrino having mass in the eV scale, the combined data from MiniBooNE and MicroBooNE still allows the 3+1 scenario. The currently running experiments ICARUS and SBND are expected to clear the smoke for this sterile neutrino situation. In future ESSnuSB and DUNE will have very strong sensitivity for the sterile neutrinos once the data from FD and ND are combined. 

Apart from the sterile neutrino, there are many other new physics scenarios which are currently being studied in the neutrino oscillation experiments. If the standard three-flavour pictures turns out to be incomplete, then one of these new physics scenarios can prove to be consistent with the observed data. Otherwise, we will get very stringent bounds on these new physics scenarios from the neutrino oscillation experiments.

\section*{Acknowledgements}

The work related to ESSnuSB is funded by the European Union. Views and opinions expressed are however those of the author only and do not necessarily reflect those of the European Union. Neither the European Union nor the granting authority can be held responsible for them. This research is also partly funded by Ministry of Science and Education of Republic of Croatia grant No. KK.01.1.1.01.0001.

\end{document}